\documentclass[osajnl,preprint,showpacs,superscriptaddress,12pt]{revtex4-1} 
\usepackage{amsmath,amssymb,graphicx}

\begin{document}

\title[]{Difficulties of the traditional PhC-based superprisms}
\author{Wei Li}
\affiliation{State Key Laboratory of Functional Materials for Informatics, Shanghai Institute
of Microsystem and Information Technology, Chinese Academy of Sciences, Shanghai 200050,
China}
\author{Xiaogang Zhang}
\email{xgzh616@gmail.com}
\affiliation{China Electronics Technology Group Corporation No.38 Research Institute, Hefei 230001, China}
\author{Xulin Lin}
\affiliation{State Key Laboratory of Functional Materials for Informatics, Shanghai Institute
of Microsystem and Information Technology, Chinese Academy of Sciences, Shanghai 200050,
China}
\author{Xunya Jiang}%
 \email{xyjiang@mit.edu}
\affiliation{Dept. of Illuminating Engineering and Light Sources, School of Information Science and Engineering, Fudan University, Shanghai 200433, China}
\affiliation{Engineering Research Center of Advanced Lighting Technology (Ministry of Education), Fudan University, Shanghai 200433, China}

\begin{abstract}
This paper is to discuss the difficulties of the traditional photonic crystals (PhC) -based superprisms.
\end{abstract}

%
\maketitle

\begin{figure*}
\centering
\includegraphics[width=0.9\columnwidth]{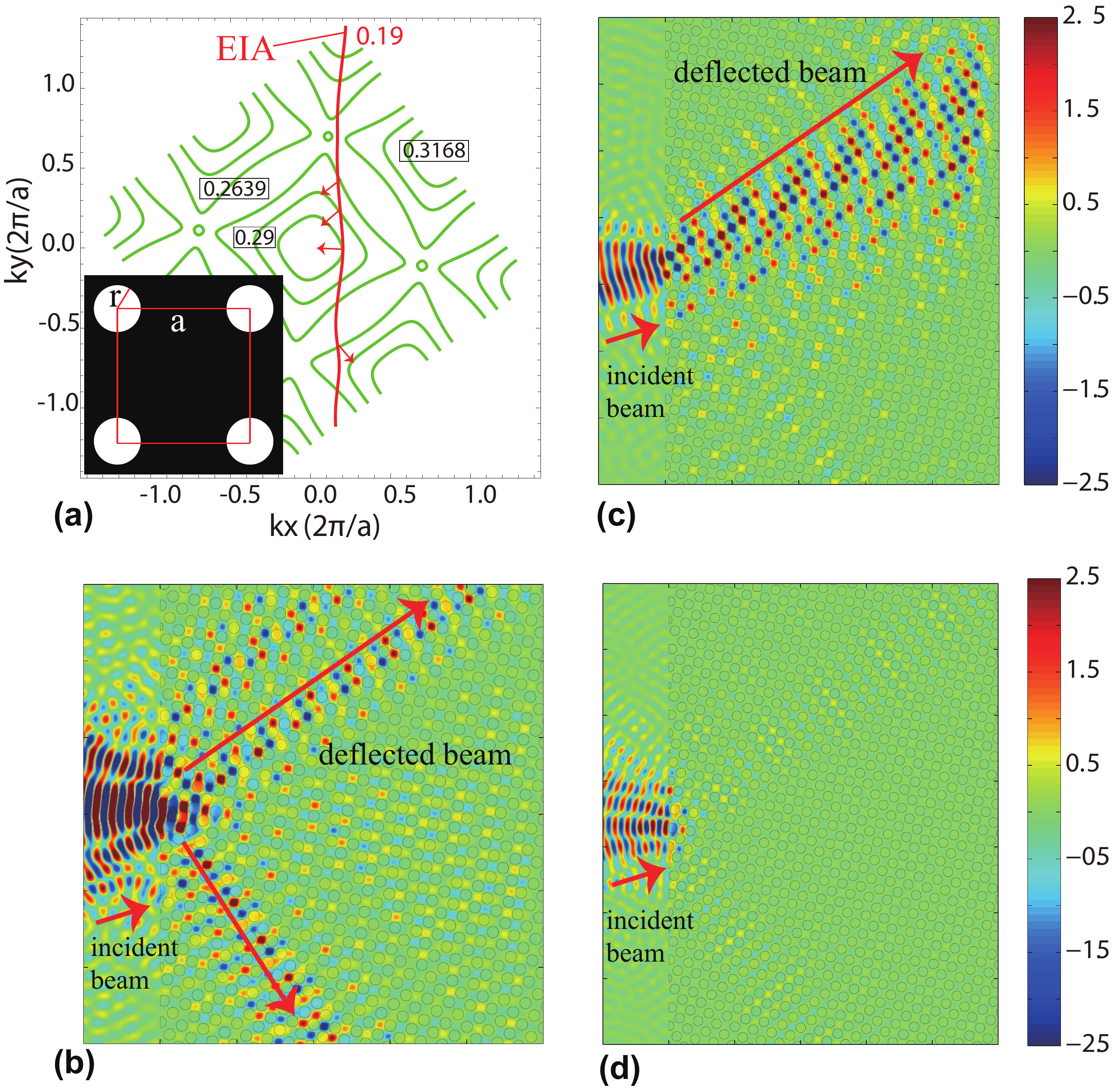}
\renewcommand\thefigure{S\arabic{figure}}
\caption{\label{fig:fig1} (a) The equi-incident angle (red line) and the EFCs (green line) of the square lattice PhC in the first BZ. The inset is the structure of the square lattice PhC. The red arrows indicate the directions of the group velocity of light at the different frequencies $\tilde{\omega}_1=0.2639$, $\tilde{\omega}_2=0.29$, and $\tilde{\omega}_3=0.3168$, respectively. (b) The irregular beam geneartion in the PhC at the frequency $\tilde{\omega}_1=0.2639$. (c) The beam divergence in the PhC at the frequency $\tilde{\omega}_2=0.29$. (d) The wavelength channel dropout at the frequency $\tilde{\omega}_3=0.3168$. The frequencies are in units of $2\pi c/a$.}
\end{figure*}

The traditional PhC-based superprisms always face three difficulties: (I) irregular beam generation, (II) beam divergence, and (III) the wavelength channel dropout in the operating frequency channel. The negative effect generated by these unfavorable factors can also be seen at Fig.4 in the reference [1], and in the reference there are also some explanations on these difficulties.

To show the difficulties of the traditional PhC-based superprisms more clearly, here we give an example. First, consider a ``sharp-corner" superprism based on a rod-type PhC made of a square lattice with the lattice constant $a$, the rod radium $r=0.35a$ and the refractive index $n=3.4$, whose structure and equi-frequency contours (EFCs) are shown in Fig.\ref{fig:fig1}(a). The best wavelength-sensitivity effect of the sharp-corner superprism can be found around the sharp-corner region of the EFCs. In order to utilize the superprism, we calculate the equi-incident angle (EIA), as shown in Fig.\ref{fig:fig1}(a) indicated by a red line. The EIA line and the high resolution of the superprism are matched together by rotating the PhC $45^\circ$. The directions of the group velocity of light in the PhC at different frequencies are labeled by red arrows, as shown in Fig.\ref{fig:fig1}(a). Second, we show the three difficulties of the superprism. Figures \ref{fig:fig1}(b), (c) and (d) show the the field distributions, which correspond to the irregular beam generation, the beam divergence, and the wavelength channel dropout, respectively. From Fig.\ref{fig:fig1}(b) to (d), the corresponding operating frequencies are $\tilde{\omega}_1=0.2639$, $\tilde{\omega}_2=0.29$, and $\tilde{\omega}_2=0.3168$, respectively. Here all the frequencies are in units of $2\pi c/a$, and all the field distributions are calculated by the finite-difference time-domain (FDTD) method.

When the operating frequency is close to the edge of Brillouin Zone (BZ), the incident light beam from homogenous media may much more easily couple to multi-modes of the PhC in higher order BZ, besides the one in the 1st BZ. In this case, the multi-refraction problem can be generated. The irregular beams always appear due to the multiply cross points induced by the parallel component of the wavevector, as shown in Fig.\ref{fig:fig1}(a). Although in some special cases the irregular beam can be small\cite{IEEE38}, we notice that, generally the intensities of irregular beams are out of control. Physically, in order to suppress the irregular beams generation one possible solution is to choose the operating region far away from the edge of BZ.

When the operating frequency has a big curvature iso-frequency contour in the BZ, for instance, $\tilde{\omega}_2=0.29$ in Fig.\ref{fig:fig1}(a), the deflected beams in the PhCs would be broaden. Unfortunately, the sharp-corner superprism always have operating frequencies near the sharp-corner region whose iso-frequency contours must have big curvature. Therefore the beam divergence is unavoidable.

When the operating frequency is at the corner of EFCs,  such as the operating frequency is $\tilde{\omega}_2=0.3168$ in Fig.\ref{fig:fig1}(a), the light cannot propagate in the PhC, as shown in Fig.\ref{fig:fig1}(d). This is because the operating frequency at the cross point cannot clearly define the wavevector. This fact makes the operating frequency cannot be switched successively, i.e., some of the wavelength channel is lost\cite{IEEE38}. We call this wavelength channel loss as ``wavelength channel dropout".

\nocite{*}

\begin{thebibliography}{s1}

\bibitem{IEEE38} T. Baba, and  M. Nakamura ,  IEEE Journal of quantum electronics,  \textbf{38}, 909(2002)


\end{thebibliography}

\end{document}